\documentclass[preprint,preprintnumbers,amsmath,amssymb,12pt,floatfix,epsfig,nofootinbib,superscriptaddress]{revtex4}
\usepackage[usenames, dvipsnames]{color}
\usepackage{graphicx}
\usepackage{amssymb}
\usepackage[normalem]{ulem}
\renewcommand\sout{\bgroup\color[rgb]{0,0,1} \ULdepth=-.5ex \ULset}

\begin{document}
\title{Effect of neutrino electromagnetic properties on the quasielastic neutral-current neutrino-nucleus scattering}
\author{K. S. Kim}
\email{kyungsik@kau.ac.kr}
\affiliation{School of Liberal Arts and Science, Korea Aerospace University, Goyang 10540, Korea}
\author{P. T. P. Hutauruk}\email{phutauruk@pknu.ac.kr}
\author{Seung-il Nam}\email{sinam@pknu.ac.kr}
\affiliation{Department of Physics, Pukyong National University, Busan 48513, Korea}
\author{Chang Ho Hyun}
\email{hch@daegu.ac.kr}
\affiliation{Department of Physics Education, Daegu University, Gyeongsan 38453, Korea}

\date{\today}
\begin{abstract}
In the quasielastic region, we investigate the effect of neutrino electromagnetic properties constrained from the recent experiments 
on the electroweak neutral current reaction process of the neutrino-$^{12}$C scattering. 
For a relativistic description of the nuclear dynamics, we employ the relativistic mean-field model, 
which has been proven to describe the data nicely in the quasielastic region.
In the present work, we analyze the influence beyond the Standard Model by considering 
the neutrino magnetic and electric dipole form factors 
and charge radius on the neutrino electroweak interactions within $^{12}$C. 
To this end, we use the values of the neutrino charge radius and the magnetic moment at the 
squared four momentum transfer $Q^2=0$ obtained from the recent experiments and
calculate the neutrino differential cross section of the neutrino-$^{12}$C scattering.
We find that the effect of the charge radius and the electric dipole form factor is very small, 
but the role of the magnetic dipole form factor is sensitive to $Q^2$ and becomes sizable at small momentum transfer.
\end{abstract}
\maketitle

\section{Introduction}

Exploring the physical properties of elementary particles is the most fundamental scientific challenge
either theoretically or experimentally.
Neutrino is one of the interesting particles in that even though about 100 years have passed 
since it was postulated theoretically by Enrico Fermi in the double beta decay reaction process, 
its physical properties, such as mass, electromagnetic properties, and their physical impact on the Universe, are still ambiguous.
For instance, it has been one of the candidates for dark matter, 
but its mass is not yet determined accurately, and flavor mixing is still an open question.
In the Standard Model (SM), it is widely known that neutrinos are assumed to be point particles, 
meaning that they don't have internal structures~\cite{sm1973}.
However, it has long been a challenging issue to pin down the limit of the SM assumption.
Recently, some evidence from either theories or experiments~\cite{pan2009, prd2022} indicates that neutrino has 
internal structures such as the magnetic moment and charge radius that the SM could not accommodate.

The charge radius and magnetic moment, which are the electromagnetic (EM) properties of the neutrinos open a way to verify the SM.
Experimentally, there has been a long history of determining the upper limits of these quantities from various experiments 
by using accelerators~\cite{acc1, acc2, acc3, acc4}, reactors~\cite{reac1, reac2} and the neutrinos coming from the sun to earth 
by using modern telescopes~\cite{borexino, xenonnt}, 
but the experimental results are still far from the converging conclusions.
In Ref. \cite{rmp2015}, several experiments and the current progress have been comprehensively discussed concerning the determination 
of the upper limit of the neutrino magnetic moment from the neutrino scattering off electrons.
However, the upper limit of the neutrino magnetic moment is uncertain by orders of magnitude in the range 
$(10^{-11}$--$10^{-7}) \mu_{\rm B}$, where $\mu_{\rm B}$ is the Bohr magneton.
Additionally, most data indicates that the charge radius is less uncertain than the magnetic moment with the order of $10^{-3}$ fm, 
but the magnitude differs by a few factors~\cite{prd2018}.
Recently, better constraints on the neutrino EM properties have been measured in the most updated experiment of 
coherent elastic neutrino-nucleus scattering (CE$\nu$NS).
It is worth mentioning that results reported up to now are still close to the limits obtained from other experiments \cite{jhep2022}, 
meaning that there is only minimal improvement from the current experiment.

Quasielastic (QE) neutrino-nucleus scattering is a useful tool to understand the nuclear structure and dynamics of the nucleons in the nuclear medium.
Accurate measurements accumulated from experiments such as MiniBooNE~\cite{mini1,mini2,mini3,mini4}, 
MicroBooNE~\cite{microboone1,microboone2}, SciBooNE~\cite{sciboone1,sciboone2} and 
T2K~\cite{t2k1,t2k2} Collaborations require precise descriptions of the wave functions of the in-medium nucleons and their interactions with the neutrino.
In recent works, the effect of the nuclear structure has been investigated thoroughly by considering the uncertainties arising from 
the in-medium effective mass of the nucleon and the density dependence of the symmetry energy \cite{prc2021,prc2022,plb2022}.
It is shown that the influence of the effective mass can be distinguished clearly, and the agreement with the experimental data is in favor of 
the isoscalar effective mass close to the nucleon mass in free space.
Another critical uncertainty from the properties of the nucleon in the nuclear medium is the magnitude of the axial mass.
With the standard value $M_{\rm A}=1.032$ GeV, theoretical results of the QE neutral-current (NC) scattering of both neutrinos 
and anti-neutrinos are consistently smaller than the measured data \cite{prc2015}.
To overcome this problem, it is required to consider meson-exchange (MEC), 
multi-nucleon processes like 2p-2h excitation, 
and final state interaction (FSI) to be able to describe the data as discussed in Refs.~\cite{prc2013,sobczyk}.
However, since the contribution of the neutrino EM properties is not affected by the inelastic channels, 
we focus on the neutrino magnetic moment and charge radius in the electroweak 
neutral-current reaction process of the neutrino scattering off the nucleus to see their effect.

In this work, the charge radius and magnetic moment of the neutrino are considered in the QE electroweak NC scattering 
of the neutrinos with $^{12}$C.
The wave function of the target nucleus is described in terms of the quantum hadrodynamics (QHD) model, 
which is proven to reproduce the electron-nucleus and neutrino-nucleus date accurately in the QE region.
Ambiguity due to the axial mass is avoided by using only the standard mass $M_{\rm A}=1.032$ GeV.
The upper limit of the magnetic moment $\mu_\nu$ differs by an order of $10^4$ maximally, but the recent measurements 
using the reactor anti-neutrinos do not exceed $10^{-10} \mu_{\rm B}$.
In our calculation, we consider two values, $\mu_{\nu_e} \simeq 2.13 \times 10^{-10} \mu_{\rm B}$ by the 
CE$\nu$NS Collaboration \cite{jhep2022}, and the upper limit of $\mu_{\nu_e} < 2.9 \times 10^{-11} \mu_{\rm B}$ 
by the GEMMA experiment \cite{gemma}.
The value of GEMMA is the smallest upper limit determined so far.
For the charge radius, we also adopt two values, $R^2_{\mu_e} = 6.42 \times 10^{-11}$ MeV$^{-2}$ from the 
CE$\nu$NS Collaboration \cite{jhep2022}, and $R^2_{\mu_e} = 6.15 \times 10^{-12}$ MeV$^{-2}$ from the BNL Collaboration \cite{acc1}.
By including the neutrino EM properties in the scattering cross section,
we expect that we can see how large their contribution is to the cross section of the QE process.
More interestingly, in comparison with the experimental data~\cite{jhep2022,gemma,acc1}, 
our result can provide constraints on the neutrino EM properties by adjusting the theoretical results to the experimental data. 
In other words, the neutrino with charge radius and magnetic moment could be an alternative to 
reduce the gap between experiment data and theory prediction 
in the QE neutrino-nucleus scattering.

The outline of this paper is organized as follows.
In Sec. II, we briefly describe our theoretical formalism for the electroweak NC reaction process of the neutrino-$^{12}$C scattering in the framework of the relativistic mean-field model \cite{horowitz} by accounting for the neutrino charge radius and magnetic moment.
Section III presents the numerical results and discussions on the differential cross sections for various neutrino magnetic moments 
and charge radii constrained by recent experiments. Finally, we summarize our work in Sec. IV.

\section{formalism}
In this section, we present the theoretical framework for the electroweak NC neutrino scattering off the nucleus 
within the QHD model in the QE regime, including the neutrino magnetic moment and charge radius. 
In the QE inclusive electroweak NC neutrino-nucleus scattering, we choose a coordinate system in which the target nucleus is placed at the origin.
The four momenta of the particles involved in the process are denoted as 
$p_i^{\mu}=(E_i, {\bf p}_i)$, $p_f^{\mu}=(E_f, {\bf p}_f)$, $p_A^{\mu}=(E_A, {\bf p}_A)$, $p_{A-1}^{\mu}=(E_{A-1}, {\bf p}_{A-1})$, 
and $p^{\mu}=(E_N, {\bf p})$ for the incident neutrino, the outgoing lepton, the target nucleus, the residual nucleus, 
and the knocked-out nucleon, respectively. 
We briefly introduce the general expression of the fivefold differential cross section for the electroweak NC neutrino scattering 
off the nucleus (in this work $^{12}$C):
\begin{eqnarray}
    \label{cs-0}
    \frac{d^5 \sigma}{dT_N d^2 \Omega_l d^2 \Omega_N} &=& \frac{M_N M_{A-1}}{(2\pi)^5 M_A} {{\bf p}_f^2 |{\bf p}|}f_{\rm{rec}}^{-1} \overline{\sum_{i,f}} \big|M_{fi}\big|^2,
\end{eqnarray}
where $M_A$, $M_{A-1}$, and $M_N$ are the nucleus, residual nucleus, and nucleon masses, respectively. 
The $\Omega_l$ and $\Omega_N$ are, respectively, the scattering directions of the outgoing lepton and the outgoing nucleon. 
The recoil factor $f_{\rm rec}$ is written as
\begin{equation}
f_{\rm rec} = {\frac {E_{A-1}} {M_A}} \left | 1 + {\frac {E_N}
{E_{A-1}}} \left [ 1 - {\frac {{\bf p} \cdot {\bf q}} {p^2}}
\right ] \right |,\end{equation}
where $\mathbf{q}=\mathbf{p}_f-\mathbf{p}_i$.

Considering the EM properties of the neutrino in the electroweak NC scattering, the squared invariant matrix element is given by 
summing the weak, electromagnetic, and interference contributions
\begin{eqnarray}
\overline{\sum_{i,f}} \big|M_{fi}\big|^2 &=&  \left[L^{\mu\nu} W_{\mu \nu} \right]_{\rm{W}} +  \left[L^{\mu\nu} W_{\mu \nu} \right]_{\rm{EM}} + \left[L^{\mu\nu} W_{\mu \nu} \right]_{\rm{INT}},
\end{eqnarray}
where the subscripts of $\rm{W}$, $\rm{EM}$, and $\rm{INT}$ represent the weak, electromagnetic, and interference 
 terms, respectively. For a free nucleon, the NC operator comprises the weak, EM, axial-vector, and pseudoscalar form factors:
\begin{eqnarray}
{\hat {\bf J}}^{\mu}_{\rm{W}} &=& F_{1}^V (Q^2){\gamma}^{\mu}
+{\frac {i F_{2}^V(Q^2)} {2M_N}}{\sigma}^{\mu\nu}q_{\nu} + G_A(Q^2)
\gamma^{\mu} \gamma^5 + {\frac {G_P(Q^2) } {2M_N}}q^{\mu}
\gamma^5, \\
{\hat {\bf J}}^{\mu}_{\rm{EM}} &=& F_{1}^{\rm{EM}} (Q^2){\gamma}^{\mu}+ 
{\frac {iF_{2}^{\rm{EM}}(Q^2)} {2M_N}}{\sigma}^{\mu\nu}q_{\nu},
\label{relJ}
\end{eqnarray}
where $F_1^{\rm{EM}} =1$ and $F_2^{\rm{EM}} = \kappa_p$ for the proton, $F_1^{\rm{EM}} =0$ and 
$F_2^{\rm{EM}}= \kappa_n$ for the neutron, and the squared four momentum transfer $Q^2 = \mathbf{q}^2 -\omega^2$.
$\kappa_n = -1.913$ and $\kappa_p = 1.793$ in the units of the nuclear magneton $\mu_N = e/2M_N$. 
By the conservation of the vector current (CVC) hypothesis with the inclusion of an isoscalar strange quark contribution, $F_i^s
(Q^2)$, the vector form factors for protons and neutrons,
$F_{i}^{V,~p(n)} (Q^2)$, are expressed as
\begin{eqnarray}
F_i^{V,~ p(n)} ( Q^2) &=&\left ( {\frac 1 2} - 2 \sin^2 \theta_W \right )
F_i^{p(n)} ( Q^2) - {\frac 1 2} F_i^{n(p)}( Q^2) -{\frac 1 2}
F_i^s ( Q^2)~ , \label{vector}
\end{eqnarray}
where $\theta_W$ is the Weinberg angle given by $\sin^2 \theta_W = 0.224$. The strange vector form factors are given as
\begin{equation}
F_1^s(Q^2) = {\frac {F_1^s (0) Q^2} {(1+\tau)(1+Q^2/M_V^2)^2}}~,
\;\;\;\;\; F_2^s(Q^2) = {\frac {F_2^s(0)}
{(1+\tau)(1+Q^2/M_V^2)^2}}~,
\end{equation}
where $\tau=Q^2/(4M_N^2)$, $M_V=0.843$ GeV,
$F_1^s (0) =-\langle r_s^2\rangle/6=0.53$ GeV$^{-2}$, and $F_2^s(0)=\mu_s$ with
the strange magnetic moment given by $\mu_s=-0.4$. 
The axial form factor is given as
\begin{eqnarray}
G_A (Q^2) &=&{\frac 1 2} \frac{(\mp g_A + g_A^s)}{(1+Q^2/M_A^2)^2},
\label{gs}
\end{eqnarray}
with $g_A=1.262$ and $M_A=1.032$ GeV. The $g_A^s=-0.19$ represents
the strange quark contents in the nucleon. The $(-,+)$ signs
coming from the isospin dependence denotes the final (proton, neutron), respectively. 
Note that the pseudoscalar form factor $G_P$ is negligible in this work because its contribution is proportional to $m_l^2/M_N^2$ where $m_l$ is the lepton mass in free space.

Next, the operators for the neutrino weak and EM currents \cite{emform1,emform2,emform3,emform4} 
are respectively given by
\begin{eqnarray}
\hat{\mathbf{j}}^{\mu}_{\rm{W}} &=& \gamma^{\mu} ( 1 - \gamma^5 ) , \\
\hat{\mathbf{j}}^{\mu}_{\rm{EM}} &=& f_m \gamma^{\mu} + g_1 \gamma^{\mu} \gamma^5 - ( f_2 + i g_2 \gamma^5 )
\, {\frac {(p_i^{\mu}+p_f^\mu)} {2m_e}},
\end{eqnarray}
where $f_m$ is written as $f_m = f_1 + (m_{\nu}/m_e) f_2$, and $m_{\nu}$ and $m_e$ denote the neutrino and electron masses, respectively.
$f_1$, $g_1$, $f_2$, and $g_2$ are the Dirac, anapole, magnetic, and electric form factors of the neutrino, respectively.
The Dirac and anapole form factors are related to the vector and axial vector charge radii $\langle R^2_V \rangle$ and $\langle R^2_A \rangle$ through \cite{emform1}
\begin{eqnarray}
f_1 (Q^2) = {\frac 1 6} \langle R^2_V \rangle Q^2, \nonumber \\
g_1 (Q^2) = {\frac 1 6} \langle R^2_A \rangle Q^2.
\end{eqnarray}
The neutrino charge radius is defined by $\langle R^2 \rangle = \langle R^2_V \rangle + \langle R^2_A \rangle$ 
and then $f_m + g_1 = {\frac 1 6} \langle R^2 \rangle Q^2$.
In the limit of $Q^2 \rightarrow 0 $ (rest frame), $f_2 (0)$ and $g_2 (0)$ define the neutrino magnetic and electric dipole moment \cite{emform3}
\begin{equation}
\mu^{m}_{\nu} = f_2 (0) \mu_B  \;\;\;\;\; \mbox{and} \;\;\;\;\;  \mu^e_{\nu} = g_2 (0) \mu_B ,
\end{equation}
where $\mu^2_{\nu} = (\mu^{m}_{\nu} )^2 + (\mu^e_{\nu} )^2 $ with the Bohr magneton $\mu_B$.

The neutrino tensor of the weak term is given by
\begin{equation}
L^{\mu \nu}_{\rm{W}} = 8 \left[ p^{\mu}_i p^{\nu}_f + p^{\nu}_i p^{\mu}_f
 - g^{\mu \nu} ( p_i\cdot p_{f} ) - i \varepsilon_{\mu \rho \nu \sigma} p^{\rho}_f p^{\sigma}_i \right],
\end{equation}
for the EM term,
\begin{eqnarray}
L^{\mu \nu}_{\rm{EM}} &=& 4 ( f^2_m + g^2_1 ) \left[ p^{\mu}_i p^{\nu}_f + p^{\nu}_i p^{\mu}_f 
- g^{\mu \nu}( p_i\cdot p_{f} ) \right] 
+8i f_m g_1 \varepsilon_{\mu \rho \nu \sigma} p^{\rho}_f p^{\sigma}_i  \nonumber \\
&& + {\frac { f^2_2 + g^2_2} {m^2_e}}( p_i\cdot p_{f} ) ( 2 p^{\mu}_i p^{\nu}_f + 2 p^{\nu}_i p^{\mu}_f
+ q^{\mu} q^{\nu} ),
\end{eqnarray}
and for the interference contribution,
\begin{equation}
L^{\mu \nu}_{\rm{INT}} = 4 ( f_m + g_1 ) \left[ p^{\mu}_i p^{\nu}_f + p^{\nu}_i p^{\mu}_f
 - g^{\mu \nu} ( p_i\cdot p_{f} ) - i \varepsilon_{\mu \rho \nu \sigma} p^{\rho}_f p^{\sigma}_i \right].
\end{equation}

Within the laboratory frame, by choosing the three momentum transfer $\bf q$ along $z$-axis, 
the inclusive differential cross section of neutrino scattering off the nucleus in the electroweak NC reaction is given by 
the contraction between lepton and hadron tensors and integrating the cross section over the phase space of scattered lepton 
and outgoing nucleon:
\begin{eqnarray}
{\frac {d\sigma} {dT_N}} &=& 4\pi^2{\frac {M_N M_{A-1}} {(2\pi)^3
M_A}} \int \sin \theta_l d\theta_l \int \sin \theta_N d\theta_N |{\bf p}|
f^{-1}_{\rm rec} 
\cr
&\times&\sum_i \sigma^i_M  \big( v_L^i R_L^i  + v_T^i R_T^i + h v_T'^i R_T'^i \big),
\label{cs}
\end{eqnarray}
where $\theta_l$ is the scattering angle of the lepton, $\theta_N$ is the polar angle of outgoing nucleons, 
$T_N$ is the kinetic energy of the knocked-out nucleon, $h=-1$ corresponds to the intrinsic helicity of the 
incident neutrino, and $i$ denotes W, EM, and INT. 
For the NC reaction, the kinematic factors $\sigma^{\rm{W}}_M$, $\sigma^{\rm{EM}}_M$, 
and $\sigma^{\rm{INT}}_M$ are defined as
\begin{eqnarray}
\sigma^{\rm{W}}_M &=& \left[ {\frac {G_F \cos\left(\frac{\theta_l}{2}\right) E_f M_Z^2}
{{\sqrt 2} \pi (Q^2 + M^2_Z)}} \right]^2, \\
\sigma^{\rm{EM}}_M &=& \left[ \frac{ \alpha_{EM} E_f \cos\left(\frac{\theta_l}{2}\right)}{{\sqrt 2} Q^2 } \right]^2, \\
\sigma^{\rm{INT}}_M &=& \left[\frac{ G_F \alpha_{EM} E_f^2 \cos^2\left(\frac{\theta_l}{2}\right) M_Z^2 }{2 \sqrt{2} \pi Q^2 (Q^2 + M^2_Z) }\right],
\end{eqnarray}
where $M_Z =$ 91.19 GeV is the rest mass of $Z$-boson.
$G_F =$ 1.166 $\times 10^{-5}$ GeV$^{-2}$ denotes the Fermi constant 
and $\alpha_{\rm EM}$ is the EM fine structure constant. 
The $R_L$, $R_T$, and $R^{'}_T$ are the longitudinal, the transverse, and the transverse interference response functions.
The appropriate response functions are written as
\begin{eqnarray}
R_L^i=\left | N^0_i(\mathbf{q}) - {\frac {\omega} {\mathbf{|q|}}} N^z_i(\mathbf{q}) \right |^2,
\,\,R_T^i=|N^x_i (\mathbf{q})|^2 + |N^y_i (\mathbf{q})|^2, \,\, R'^i_T = 2
{\mbox {Im}}\left[{N^x_i}^* (\mathbf{q})N^y_i (\mathbf{q})\right].
\end{eqnarray}
$N^\mu_i(\mathbf{q})$ is the Fourier transform of the nucleon transition current $J^\mu_i(\mathbf{r})$ given by
\begin{eqnarray}
N^\mu_i &=& \int J^\mu_i (\mathbf{r}) e^{i \mathbf{q} \cdot \mathbf{r}} d^3 r,\\
J^\mu_i(\mathbf{r}) &=& e \bar{\psi}_N \hat{\mathbf{J}}^\mu_i \psi_b(\mathbf{r}),
\end{eqnarray}
where $\psi_N$ and $\psi_b$ are the wave functions of the outgoing and the bound state nucleons, respectively.

The neutrino kinematics factors for the weak term are given by
\begin{eqnarray}
v^W_L &=& 1, \\
v^W_T &=& \tan^2 {\frac {\theta_l} {2}} + {\frac {Q^2} {2q^2}}, \\
{v' }^W_T &=& \tan^2 {\frac {\theta_l} {2}} {\sqrt {\tan^2 {\frac {\theta_l} {2}} + {\frac {Q^2} {q^2}}}},
\end{eqnarray}
for the EM part,
\begin{eqnarray}
v^{\rm{EM}}_L &=& 2 ( f^2_m + g^2_1 ) + {\frac { f^2_2 + g^2_2} {2 m^2_e}} (p_i + p_f )^2 \tan^2 {\frac {\theta_l} {2}} \label{eq:em1},  \\
v^{\rm{EM}}_T &=& 4 ( f^2_m + g^2_1 ) \left ( \tan^2 {\frac {\theta_l} {2}} + {\frac {Q^2} {2q^2}} \right )  
+ \left ({\frac { f^2_2 + g^2_2} {m^2_e}} \right ) {\frac {Q^4} {2q^2}}, \label{eq:em2} \\
{v'}^{\rm{EM}}_{T} &=& 8 i f_m g_1 \tan^2 {\frac {\theta_l} {2}} {\sqrt {\tan^2 {\frac {\theta_l} {2}} + {\frac {Q^2} {q^2}}}}, \label{eq:em3}
\end{eqnarray}
and for the interference term,
\begin{eqnarray}
v^{\rm{INT}}_L &=& 2 ( f_m + g_1 ), \\
v^{\rm{INT}}_T &=& 4 ( f_m + g_1 ) \left ( \tan^2 {\frac {\theta_l} {2}} + {\frac {Q^2} {2q^2}} \right ), \\
{v'}^{\rm{INT}}_{T} &=& -4 i ( f_m + g_1 ) \tan^2 {\frac {\theta_l} {2}} {\sqrt {\tan^2 {\frac {\theta_l} {2}} + {\frac {Q^2} {q^2}}}}.
\end{eqnarray}

\section{Result}
Using the QHD nuclear model, we study the role of neutrino EM property.
To do this, we calculate the single differential cross sections of the NC interaction off $^{12}$C in terms of the kinetic energy of the final nucleon and the squared four momentum transfer.
We use two values of neutrino charge radius: one is determined by the combined Dresden-II and COHERENT analysis 
which is $\langle R^2 \rangle \sim 5 \times 10^{-32}$ cm$^2$ \cite{jhep2022} and the other was obtained from 
the BNL-E734, $\langle R^2 \rangle \sim 0.48 \times 10^{-32}$ cm$^2$ \cite{acc1}, which is smallest value that we found.
For the magnetic moment, we also take two values: one is $\mu_{\nu} \sim 2.13 \times 10^{-10} \mu_B$ from Dresden-II and the other is $\mu_{\nu} \sim 2.9 \times 10^{-11} \mu_B$ \cite{gemma} that is the smallest value.
Our results are presented for four cases as summarized in Tab.\,\ref{tab1}.

\begin{table}
\begin{center}
\begin{tabular}{c|c|c}
 &\,\, $\langle R^2 \rangle$ [cm$^2$]\,\, &\,\, $\mu_\nu/\mu_B$\,\, \\ \hline
case 1\,\, &\,\, $5\times 10^{-32}$\,\, &\,\,$2.13\times 10^{-10}$ \\
case 2\,\, &\,\, $0.48 \times 10^{-32}$ \,\,&\,\, $2.13\times 10^{-10}$ \\
case 3\,\, &\,\,  $5\times 10^{-32}$ \,\,&\,\, $2.90\times 10^{-11}$ \\
case 4\,\, &\,\, $0.48 \times 10^{-32}$\,\, &\,\, $2.90\times 10^{-11}$ \\
\end{tabular}
\end{center}
\caption{Charge radii and magnetic moments of the neutrino used in the calculation.}
\label{tab1}
\end{table}


\begin{figure}
\centering
\includegraphics[width=15cm]{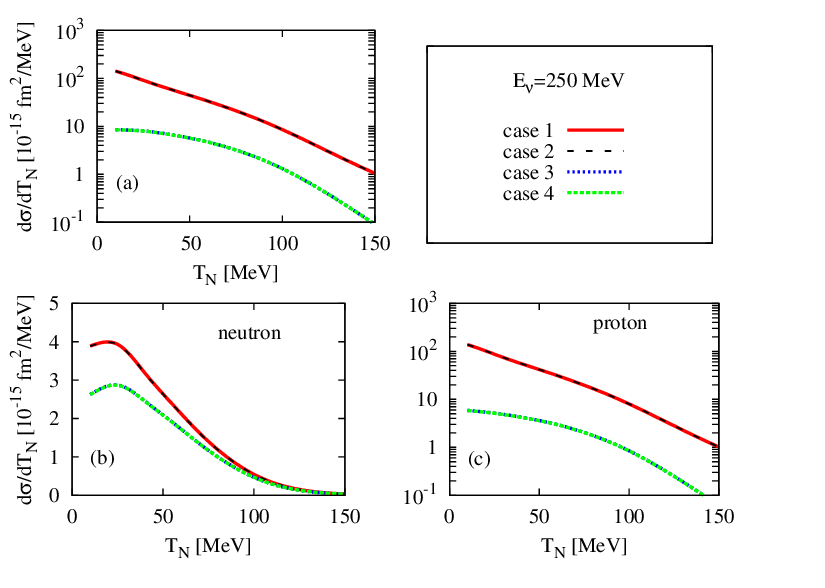}
\caption{The differential cross sections of $(\nu, \nu')$ reaction with $^{12}$C target in terms of the kinetic energy of the outgoing
nucleon and the incident neutrino energy is $E_i = 250$ MeV.
The solid (red) lines are the results for the case 1, the dashed (black) lines are for the case 2, the dotted (blue) lines are for the case 3, 
and the short-dashed (green) lines are for the case 4.
The panel (a) is the results for the sum of contributions from the proton and the neutron, 
the panel (b) is for the neutron participating in the reaction,
and the panel (c) is the result of the proton.}
\label{fig1}
\end{figure}


\begin{figure}
\centering
\includegraphics[width=15cm]{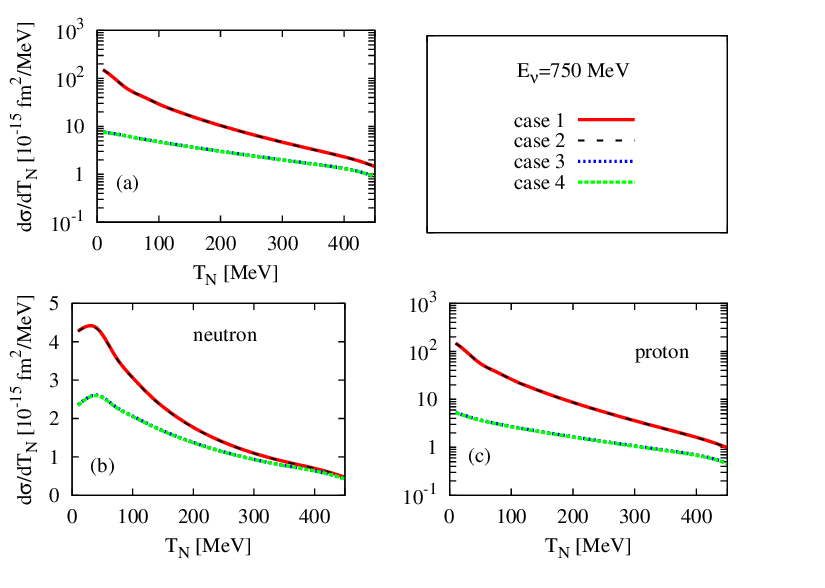}
\caption{The same as the Fig. \ref{fig1} but with the incident neutrino energy $E_i=750$ MeV.}
\label{fig2}
\end{figure}

As shown in Figs. \ref{fig1} and \ref{fig2}, the single differential cross sections are represented  in terms of the kinetic energy of the final nucleon 
at the incident neutrino energies $E_i = 250$ and 750 MeV.
The solid (red) lines are the results for case 1, the dashed (black) lines are for case 2, 
the dotted (blue) lines are for case 3, and the short-dashed (green) lines are for case 4.
The panels (b) and (c) are the cross sections for only neutron and proton participating in the reaction, 
and the (a) is the summation of (b) and (c).
The differences between the red and black curves and between the blue and green curves are very small, 
which is less than 1 \%, that is, the effect of the charge radius of the neutrino is very small.
However, the difference between the red (black) and blue (green) is very large, in particular, the deviation is mainly from the proton part.

With the small $\mu_\nu$ value ($2.9 \times 10^{-11}\mu_B$), 
contribution of the proton is similar in magnitude to that of the neutron,
but with the large $\mu_\nu$ value ($2.19 \times 10^{-10}\mu_B$), while neutron shows slight enhancement from the small $\mu_\nu$ result,
contribution of the proton increases by a factor larger than 10.
In order to understand the huge enhancement by the proton, we diagnose the contributions from weak, electromagnetic, and
interference terms.


\begin{figure}
\centering
\includegraphics[width=13cm]{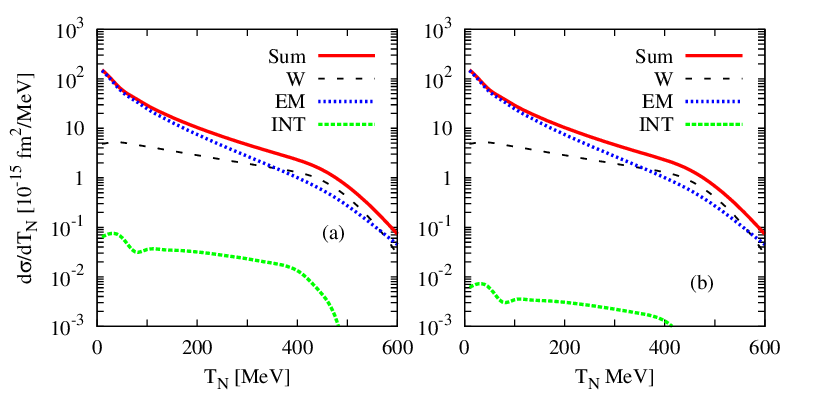}
\includegraphics[width=12.5cm]{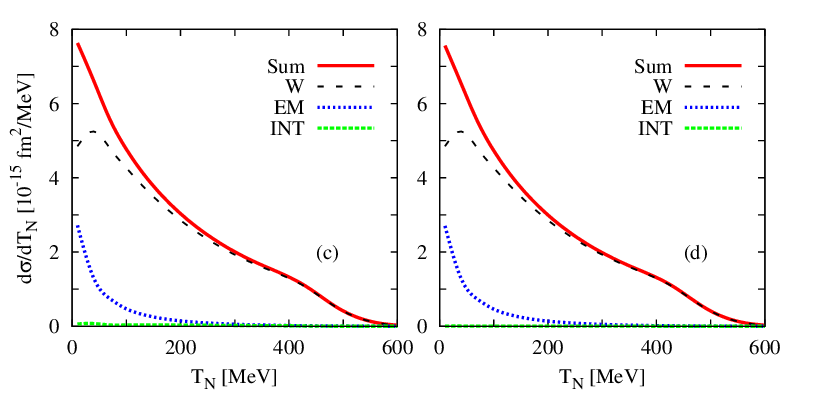}
\caption{Breakdown of the differential cross sections of $(\nu, \nu')$ reaction in terms of the kinetic energy 
of the outgoing nucleon with the incident neutrino energy $E_i = 750$ MeV.
The solid (red) lines are the results for the sum of three terms, the dashed (black) lines are for the weak interaction, 
the dotted (blue) lines are for the EM interaction, and the short-dashed (green) lines are for the interference term.
The panel (a) is the results for the case 1, (b) is for the case 2, (c) is for the case 3, and (d) is for the case 4.}
\label{fig3}
\end{figure}

In Fig. \ref{fig3}, we investigate the contribution of each term in Eq. (\ref{cs}) at the incident energy $E_i=750$ MeV.
The solid (red) lines are the results of the summing over the whole terms, the dashed (black) lines are for the weak interaction, 
the dotted (blue) lines are for the EM interaction, and the short-dashed (green) curves are for the interference term.
The panels (a), (b), (c), and (d) are for case 1, case 2, case 3, and case 4, respectively.
The EM interaction is dominant for the cases 1 and 2, but the weak interaction is dominant for cases 3 and 4.
The cross section is sensitive to the magnetic moment of neutrino but not to the charge radius like the previous results.
Nevertheless, a tiny dependence on the charge radius is found by comparing the interference terms in the left panels with 
those in the right panels.
Looking at the figures in the upper row, the contribution of the interference term in the left panel is larger by a factor of about 10 than the one in the right panel. The factor 10 is similar to the ratio of the $\langle R^2 \rangle$ values between the left and right panels.
However, the dependence on the charge radius can hardly be seen in the weak and EM terms, 
so practically no difference arises from the charge radius in the total result.

A striking result happens in the EM contribution.
With the large $\mu_\nu$ as in Figs. 3(a, b), weak term contributions are identical to those of the small $\mu_\nu$ results,
but huge enhancement of the EM contribution occurs at $T_N$ roughly below 500 MeV,
and the cross section is dominated by the EM term at $T_N \lesssim 300$ MeV.
With the small $\mu_\nu$ value, the contribution of the EM term is marginal at $T_N \gtrsim 200$~MeV,
and becomes substantial at $T_N \lesssim 100$~ MeV.
It is notable that the EM contribution increases as $T_N$ decreases, and it becomes 
about 40~\% of the weak term at energies close to zero.
As will be seen in the next figure, enhancement of the EM effect at small energies has a deep impact 
in the comparison with experimental data.

Enhancement of EM contribution can be understood from Eqs.\,(\ref{eq:em1}, \ref{eq:em2}, \ref{eq:em3}).
Dominant contribution comes for the second terms in Eqs.\,(\ref{eq:em1}, \ref{eq:em2}) because
$((p_i+p_f)/m_e)^2$ or $(Q/m_e)^2$ contain small electron mass in the denominator, so they become large at the energies 
considered in the work.
In addition, because the magnitude of $f_2$ is much bigger than that of $g_2$,
$f_2$ plays a dominant role in the EM contribution.
The $\mu_\nu$ value in case 1 and case 2 is larger than the value in case 3 and case 4 by a factor $\sim 7.3$,
and as a result, we have the enhancement of the proton cross section
due to the EM terms larger than the enhancement in the neutron cross section by a factor of about 50, which is close to $7.3^2 \sim 53$ .

\begin{figure}
\centering
\includegraphics[width=15cm]{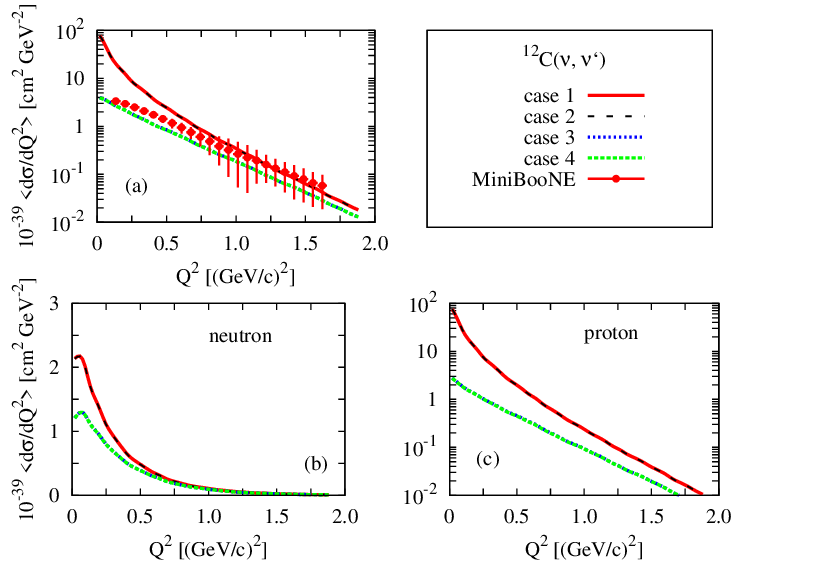}
\caption{The flux-averaged differential cross sections are compared with the experimental data measured from \cite{mini2}.
The legends of the curves are the same as Fig. \ref{fig1}.}
\label{fig4}
\end{figure}

Figure \ref{fig4} depicts the flux-averaged single differential cross section in terms of $Q^2$ by using 
four cases of the charge radius and magnetic moment like Fig.\,\ref{fig1} and shown is the comparison with the experimental data 
measured from MiniBooNE \cite{mini2}.
The panels (b) and (c) are the results for only neutron and proton participating in the reaction.
The role of the charge radius is insensitive but that of the magnetic moment is sensitive to the cross section as shown in the previous results.
In panel (a), the red and black curves
agree well with the data at $Q^2 \gtrsim 1\, ({\rm GeV}/c)^2$, but the overestimation becomes significant at 
$Q^2 \lesssim 1\, ({\rm GeV}/c)^2$.
Blue and green lines corresponding to the small value of the magnetic moment consistently underestimate the data.
The discrepancy with data is large at large $Q^2$, and becomes less as $Q^2$ decreases.

If we do not consider the neutrino EM properties, the results are the same as the weak lines in Fig~\ref{fig3}.
Large $\mu_\nu$ value is good at reproducing the data at large $Q^2$, but its diverging behavior at small $Q^2$
attaches a question mark to the large value of the magnetic moment.
With the small $\mu_\nu$ value, EM contribution is negligible at high energies but enhances the cross sections at low energies.
As a result, the cross sections are closer to the experimental data than those without the EM properties.
In the small $Q^2$ region, magnetic moment reduces the gap between theory and experiment,
and at the smallest $Q^2$ value, the difference between data and calculation almost disappears.
For the large momentum transfer, the magnetic moment is not sufficient to compensate for the discrepancy between theory and experiment,
and contributions from the inelastic processes are expected to play a dominant role in filling the gap between theory and experiment.

%
%

\section{Summary}

In the present work, we calculate the neutral-current neutrino-nucleus scattering off $^{12}$C in the quasielastic region with the relativistic nuclear model.
To guarantee current conservation and gauge invariance, the wave functions of the continuum nucleons are obtained 
by solving the Dirac equation with the same scalar and vector potentials of the bound nucleons.
Note that we do not include inelastic processes like particle-hole excitation and etc.

The single differential cross sections are calculated with four different cases of the 
neutrino charge radius and magnetic moment in terms of the kinetic energy of the knocked-out nucleon.
The influence of the charge radius is very small, and there is no practical difference in the differential cross sections.
On the other hand, the magnetic moment plays a crucial role at low energies.

The cross section of the proton is more strongly dependent on the magnitude of the magnetic moment than the neutron,
and the effect increases at low energies.
For example, with $E_i=250$ MeV, increase of the cross section from $\mu_\nu = 2.9\times 10^{-11}\mu_B$ to
$\mu_\nu = 2.13\times 10^{-10} \mu_B$ is about 50~\% for the neutron at $T_N=10$ MeV,
but the difference becomes very big for the proton at the same $T_N$.
With the $\mu_\nu = 2.9\times 10^{-11}\mu_B $, EM contribution amounts to about 60~\% of the weak one at small momentum transfer.
As a result, the EM properties of the neutrino reduce the gap between the standard result (only weak contribution) and the MiniBooNE data
substantially, and make the theory better reproduce the data at small momentum transfer.
In conclusion, with the present limits on the EM properties of the neutrino, their effect appears evidently in the QE scattering
at low energies of the outgoing nucleons. 

%

\section*{Acknowledgments}
This work was supported by the National Research Foundation of Korea (NRF) grant funded by the Korean government
(NRF-2018R1A5A1025563, 2022R1A2C1003964, 2022K2A9A1A06091761, and 2023R1A2C1003177).


\begin{thebibliography}{99}

\bibitem{sm1973}
E. S. Abers and B. W. Lee,
Phys. Rep. {\bf 9}, 1 (1973).

\bibitem{pan2009}
C. Giunti and A. Studenikin,
Phys. Atom. Nucl. {\bf 73}, 2089 (2009).

\bibitem{prd2022}
D. A. Sierra, O. G. Miranda, D. K. Papoulias, and G. S. Garcia,
Phys. Rev. D {\bf 105}, 035027 (2022).

\bibitem{acc1}
L. A. Ahrens {\it et al.}, 
Phys Rev. D {\bf 41}, 3297 (1990)

\bibitem{acc2}
R. C. Allen {\it  et al.},
Phys. Rev. D {\bf 47}, 11 (1993).

\bibitem{acc3}
CHARM-II Collaboration,
Phys. Lett. B {\bf 345}, 115 (1995).

\bibitem{acc4}
LSND Collaboration,
Phys. Rev. D {\bf 63}, 112001 (2001).

\bibitem{reac1}
G. S. Vidyakin {\it et al.},
JETP Lett. {\bf 55}, 206 (1992).

\bibitem{reac2}
TEXONO Collaboration,
Phys. Rev. D {\bf 81} 072001 (2010).

\bibitem{borexino}
S. Kumaran, L. Ludhova, \"{O}. Penek, and G. Settanta,
Universe {\bf 2021}, 231 (2021).

\bibitem{xenonnt}
XENON collaboration, 
Phys. Rev. Lett. {\bf 126}, 091301 (2021),

\bibitem{rmp2015}
C. Giunti and A. Studenikin,
Rev. Mod. Phys. {\bf 87}, 531 (2015).

\bibitem{prd2018}
M.~Cadeddu, C.~Giunti, K.~A.~Kouzakov, Y.~F.~Li, Y.~Y.~Zhang and A.~I.~Studenikin,
Phys. Rev. D \textbf{98}, no.11, 113010 (2018)
[erratum: Phys. Rev. D \textbf{101}, no.5, 059902 (2020)].


\bibitem{jhep2022}
M. A. Corona, M. Cadeddu, N. Cargioli, F. Dordei, C. Giunti, Y. F. Li,
C. A. Ternes, and Y. Y. Zhang,
JHEP {\bf 09}, 164 (2022).

\bibitem{mini1}
A. A. Aguilar-Arevalo {\it et} {\it al}.(MiniBooNE Collaboration),
Phys. Rev. D {\bf 81}, 092005 (2010).

\bibitem{mini2}A. A. Aguilar-Arevalo {\it et} {\it al}. (MiniBooNE Collaboration),
Phys. Rev. D {\bf 82}, 092005 (2010).

\bibitem{mini3}A. A. Aguilar-Arevalo {\it et} {\it al}. (MiniBooNE Collaboration),
Phys. Rev. D {\bf 88}, 032001 (2013).

\bibitem{mini4}A. A. Aguilar-Arevalo {\it et} {\it al}. (MiniBooNE Collaboration),
Phys. Rev. D {\bf 91}, 012004 (2015).

\bibitem{microboone1}P. Abratenko, {\it et} {\it al}. (MicroBooNE Collaboration), Phys. Rev. Lett. {\bf 123}, 131801 (2019); {\bf 125}, 201803 (2020); Phys. Rev. D {\bf 102}, 112013 (2020).

\bibitem{microboone2}P. Abratenko, {\it et} {\it al}. (MicroBooNE Collaboration), Phys. Rev. Lett. {\bf 128}, 151801 (2022).
    
\bibitem{sciboone1}Y. Nakajima {\it et} {\it al}. (SciBooNe Collaboration), Phys. Rev. D {\bf 83}, 012005 (2011).

\bibitem{sciboone2}K. B. M Mahn {\it et} {\it al}. (MiniBooNE and SciBooNE Collaborations), Phys. Rev. D {\bf 85}, 032007 (2012); G. Cheng {\it et} {\it al}. (MiniBooNE and SciBooNE Collaborations), Phys. Rev. D {\bf 86}, 052009 (2012).

\bibitem{t2k1}
K. Abe {\it et al.} (T2K Collaboration),
Phys. Rev. D {\bf 92}, 112003 (2015): {\it ibid}. {\bf 98}, 032003 (2018).

\bibitem{t2k2}
K. Abe {\it et al.} (T2K Collaboration),
Phys. Rev. D {\bf 93}, 112012 (2016). 

\bibitem{prc2021}
H. Gil, C. H. Hyun, and K. S. Kim,
Phys. Rev. C {\bf 104}, 044613 (2021).

\bibitem{prc2022}
H. Gil, C. H. Hyun, and K. S. Kim,
Phys. Rev. C {\bf 105}, 024607 (2022).

\bibitem{plb2022}
K. S. Kim, H. Gil, and C. H. Hyun,
Phys. Lett. B. {\bf 833}, 137273 (2022).

\bibitem{prc2015}
K. S. Kim, M.-K. Cheoun, W. Y. So, and H. Kim,
Phys. Rev. C {\bf 91}, 014606 (2015).

\bibitem{prc2013}
K. S. Kim, M.-K. Cheoun, and W. Y. So, Phys. Rev. C {\bf 88}, 044615 (2013).

\bibitem{sobczyk}
J. E. Sobczyk, J. Nieves, and F. Sanchez, Phys. Rev. C {\bf 102}, 024601 (2020).

\bibitem{gemma}
A. Beda {\it et al.},
Adv. High Energy Phys. {\bf 2012}, 350150 (2012).

\bibitem{horowitz}
C. J. Horowitz and B. D. Serot, 
Nucl. Phys. A {\bf 368}, 503 (1981).

\bibitem{emform1}
B. K. Kerimov, M. Ya Safin, and H. Nazih, 
Izy. Akad. Nauk. SSSR. Fiz {\bf 52}, 136 (1998).

\bibitem{emform2}
A. M. Mourao, J. Pulido, and J. P. Ralston, 
Phys. Lett. B {\bf 285}, 364 (1992).

\bibitem{emform3}
E. Nardi, 
AIP Conf. Proc. {\bf 670}, 118 (2003); 
M. Hirsch, E. Nardi, and D. Restrepo, 
Phys. Rev. D {\bf 67}, 033005 (2003).

\bibitem{emform4}
A. Sulaksono, C. K. Williams, P. T. P. Hutauruk, and T. Mart, 
Phys. Rev. C {\bf 73}, 025803 (2006).


%
%
%
%
%
%
%


\end{thebibliography}
\end{document}